\documentclass[a4paper]{article}
\usepackage{amsmath}
\usepackage{epsfig}
\usepackage{graphicx}  
\usepackage{color}     
\usepackage{lscape}
\usepackage{color}
\date{\today}
\begin{document}
\title{\Large \bf Broken $S_3$ Neutrinos }
{\small
\author{H. B. Benaoum \\
Prince Mohammad Bin Fahd University, Al-Khobar 31952, Saudi Arabia  \\
Email: hbenaoum@pmu.edu.sa}
}
\maketitle
\begin{center}
\small{\bf Abstract}\\[3mm]
\end{center}
Motivated by recent measurements which strongly support a nonzero reactor mixing angle $\theta_{13}$, we study a deviation from $S_3$ neutrino discrete symmetry by explicitly breaking the neutrino mass matrix with a general retrocirculant matrix. 
We show that nonzero $\theta_{13}$ and nonzero CP violation parameter $J_{CP}$ arise due to the difference between $y_2$ and $y_3$. We demonstrate that it is possible to obtain the experimentally favored results for neutrino masses and mixing angles from this mass matrix. Furthermore, we estimate the effective masses $m_{\beta}$ and $m_{\beta \beta}$ and total neutrino mass $\sum |m_i|$ predicted by this mass matrix.
\\\\
{\bf Keywords}: Neutrino Physics; Flavor Symmetry; Neutrino Mass and Mixing; Discrete Symmetries.
\\
{\bf PACS numbers}: 14.60.Pq; 12.15.Ff
\begin{minipage}[h]{14.0cm}
\end{minipage}
\vskip 0.3cm \hrule \vskip 0.5cm


Experiments using solar, atmospheric and reactor neutrinos have made considerable progress in establishing two different mass squared differences ( $\Delta m^2_{21}$ 
and $\Delta m^2_{31}$ ) and two large mixing angles ( $\theta_{12}$ and $\theta_{23}$ ) in the lepton sector. Recently MINOS ~\cite{minos1,minos2}, T2K ~\cite{t2k}, 
Double CHOOZ ~\cite{chooz}, Daya Bay ~\cite{daya} and RENO ~\cite{reno} have revealed that the reactor mixing angle $\theta_{13}$ is not only nonzero but relatively large. \\

The phenomenon of neutrino mixing can be simply described by the Pontecorvo-Maki-Nakagawa-Sakata 
( PMNS ) neutrino mixing matrix $V_{PMNS}$ \cite{maki}, which links 
the neutrino flavor eigenstates $\nu_e, \nu_{\mu}, \nu_{\tau}$ to the mass eigenstates $\nu_1, \nu_2, \nu_2$ : \\

\begin{eqnarray} 
V  & = & \left(  \begin{array}{ccc} 
V_{e 1} & V_{e 2} & V_{e 3} \\
V_{\mu 1} & V_{\mu 2} & V_{\mu 3} \\
V_{\tau 1} & V_{\tau 2} & V_{\tau 3} \\ 
\end{array}
\right) ~~~~~.
\end{eqnarray}
In the standard parametrization used by the Particle Data Group ( PDG ), the PMSN matrix is expressed by three mixing angles $\theta_{12}, \theta_{23}$ and $\theta_{13}$ and one intrinsic CP violating phase $\delta$ for Dirac neutrinos,  
\begin{eqnarray} 
V  & = &  {  \left( 
\begin{array}{ccc}
c_{12} c_{13} & s_{12} c_{13} & s_{13} e^{- i \delta} \\
- s_{12} c_{23} - c_{12} s_{23} s_{13} e^{i \delta} & c_{12} c_{23} - s_{12} s_{23} s_{13} e^{i \delta} & s_{23} c_{13} \\
s_{12} s_{23} - c_{12} c_{23} s_{13} e^{i \delta} & - c_{12} s_{23} - s_{12} c_{23} s_{13} e^{i \delta} & c_{23} c_{13} \\  
\end{array} 
\right) } . P_{Maj}
\end{eqnarray}
where $c_{ij} = \cos \theta_{ij}, s_{ij} = \sin \theta_{ij}$ and $P_{Maj}$ is a diagonal matrix with Majorana CP violating phases. \\

The three mixing angles are related to the moduli of the elements of the PMNS mixing matrix as : 
\begin{eqnarray}
\sin^2 \theta_{13} & = & | V_{e 3} |^2 \nonumber \\
\sin^2 \theta_{12} & = & \frac{| V_{e 2} |^2}{1 - | V_{e 3} |^2} \nonumber \\
\sin^2 \theta_{23} & = & \frac{| V_{\mu 3} |^2}{1 - | V_{e 3} |^2}  ~~~~~~~.
\end{eqnarray} 

The well-known tri-bimaximal (TBM) mixing pattern, which corresponds to $\theta_{13} = 0$, $\theta_{23} = \pm \frac{\pi}{4}$ and $\theta_{12} = \sin^{-1} \left( \frac{1}{\sqrt{3}} \right)$, has attracted a degree of attention in the literature because it suggests some underlying flavor symmetry among lepton's generations. 
This flavor symmetry is expected to explain the mass spectrum and neutrino mixing pattern. \\

The TBM form is : 
\begin{eqnarray}
V_0 & = & \left( \begin{array}{ccc} 
\sqrt{\frac{2}{3}} & \frac{1}{\sqrt{3}} & 0 \\
- \frac{1}{\sqrt{6}} & \frac{1}{\sqrt{3}} & - \frac{1}{\sqrt{2}} \\
-\frac{1}{\sqrt{6}} & \frac{1}{\sqrt{3}} & \frac{1}{\sqrt{2}} \end{array} \right) ~~~~.
\end{eqnarray}

Models based on discrete symmetries were successful in reproducing this matrix. Among a number of interesting discrete flavor symmetries discussed in the 
literature, the $S_3$ symmetry which is the permutation group of three objects, is the simplest \cite{wolfenstein}. $S_3$ is the smallest non-Abelian discrete group. \\
The three-dimensional reducible representations of all $S_3$ group elements are : 
\begin{eqnarray}
S^{(1)} & = & \left( \begin{array}{ccc} 
1 & 0 & 0 \\
0 & 1 & 0 \\
0 & 0 & 1 \end{array} \right) ~~~;~~~ 
S^{(12)} ~~=~~ \left( \begin{array}{ccc} 
0 & 1 & 0 \\
1 & 0 & 0 \\
0 & 0 & 1 \end{array} \right)~~~;~~~
S^{(13)} ~~= ~~\left( \begin{array}{ccc} 
0 & 0 & 1 \\
0 & 1 & 0 \\
1 & 0 & 0 \end{array} \right) \nonumber \\
S^{(23)} & = & \left( \begin{array}{ccc} 
1 & 0 & 0 \\
0 & 0 & 1 \\
0 & 1 & 0 \end{array} \right)~~~;~~~
S^{(123)} ~~=~~ \left( \begin{array}{ccc} 
0 & 0 & 1 \\
1 & 0 & 0 \\
0 & 1 & 0 \end{array} \right)~~~;~~~
S^{(132)} ~~= ~~\left( \begin{array}{ccc} 
0 & 1 & 0 \\
0 & 0 & 1 \\
1 & 0 & 0 \end{array} \right) ~~~.
\end{eqnarray}
The most general neutrino mass matrix $M_{\nu}^0$ invariant under $S_3$ is : 
\begin{eqnarray}
M_{\nu}^0 & = & \alpha ~S^{(1)} + \beta \left( S^{(12)} + S^{(13)} + S^{(23)} \right)
\end{eqnarray}
where $\alpha$ and $\beta$ are, in general, complex numbers. \\

In the basis where the charged lepton mass matrix is diagonal, the TBM mixing matrix diagonalizes the neutrino matrix $M_{\nu}^0$. 
\begin{eqnarray}
V_0^{T} M_{\nu}^0 V_0 & = & \left( \begin{array}{ccc} 
\alpha & 0 & 0 \\
0 & \alpha + \beta & 0 \\
0 & 0 & \alpha \end{array} \right) ~~~~~~.
\end{eqnarray}
This matrix leads to two degenerate masses, namely $m_1$ and $m_3$. However, this is not correct experimentally. To overcome this problem, it was suggested in \cite{jora} 
that in fact the three masses are degenerate by letting the complex number $\alpha$ lies in the third quadrant,
\begin{eqnarray}
\alpha & = & - i |\alpha| e^{-i \frac{\psi}{2}} ~~~~ \mbox{for}~~0 \leq \psi < \pi
\end{eqnarray}
and taking $\beta$ as real number such that : 
\begin{eqnarray}
\beta & = & \frac{2}{3} |\alpha| \sin \frac{\psi}{2} ~~~~~~~.
\end{eqnarray}
In this work, we consider the neutrino matrix $M_{\nu}^0$, which is invariant under $S_3$, as zeroth order with degenerate masses and TBM mixing angles. 
Nondegenerate mass spectrum and nonzero $\theta_{13}$ were realized in \cite{jora}-\cite{dev} by introducing small perturbations that violate $S_3$ symmetry. \\

Here, we investigate the phenomenological consequences of the deviation from an exact $S_3$ symmetry by explicitly breaking the neutrino mass matrix $M_{\nu}^0$ 
with a general retrocirculant matrix :
\begin{eqnarray}
\Delta M_{\nu} & = & - \alpha~ \left( \begin{array}{ccc}
y_1 &  y_2 & y_3\\
y_2 &  y_3 & y_1 \\
y_3 &  y_1 & y_2 \end{array} \right)
\end{eqnarray}
where the dimensionless parameters $y_i = |y_i| e^{i \varphi_i/2}$ are complex numbers with magnitude less than one and $0 \leq \varphi_i \leq \pi$. 
We also consider the charged lepton to be diagonal so the leptonic mixing solely comes from the neutrino sector. 
It is easy to see that $\Delta M_{\nu}$ can be written as a linear combination of $S^{(23)}$, $S^{(12)}$ and $S^{(13)}$ as :
\begin{eqnarray}
\Delta M_{\nu} & = & - \alpha \left( y_1 ~S^{(23)} +  y_2 ~S^{(12)} + y_3 ~S^{(13)} \right) ~~~~~~.
\end{eqnarray}

As a result the broken neutrino matrix $M_{\nu}$ becomes : 
\begin{eqnarray}
M_{\nu} & = & M_{\nu}^0 + \Delta M_{\nu} \nonumber \\
& = & \left( \begin{array}{ccc} 
\alpha + \beta - \alpha~ y_1 & \beta - \alpha~ y_2 & \beta - \alpha~ y_3 \\
\beta - \alpha~ y_2 & \alpha + \beta - \alpha~ y_3 & \beta - \alpha~ y_1 \\
\beta - \alpha~ y_3  & \beta - \alpha~ y_1  & \alpha + \beta - \alpha~ y_2 \end{array} \right) ~~~~ .
\end{eqnarray}
The eigenvalues of the above  matrix are : 
\begin{eqnarray}
m_1 & = & \alpha - \alpha~ \sqrt{y_1^2 + y_2^2 + y_3^2 - y_1 y_2 - y_1 y_3 - y_2 y_3} \nonumber \\
m_2 & = & \alpha + 3 \beta - \alpha ~\left( y_1 + y_2 + y_3 \right) \nonumber \\
m_3 & = & \alpha + \alpha~ \sqrt{y_1^2 + y_2^2 + y_3^2 - y_1 y_2 - y_1 y_3 - y_2 y_3}~~~~~~.
\end{eqnarray}

The matrix $M_{\nu}$ is called magic mass matrix since every row and column add up to the same value which is $m_2$ for this case. 
It implies that this mass matrix has a trimaximal eigenvector $\left( \frac{1}{\sqrt{3}}, \frac{1}{\sqrt{3}},  \frac{1}{\sqrt{3}} \right)^T$ \cite{kumar}. Moreover the 
$\mu-\tau$ symmetry corresponding to $\theta_{13} = 0$ and $\theta_{23} = \pm \frac{\pi}{4}$ is broken for $M_{\nu}$,
\begin{eqnarray}
[M_{\nu}, S^{(23)}] & = & - \alpha~ (y_2 - y_3) ~\left( S^{(123)} - S^{(132)} \right) 
\end{eqnarray} 
due to the difference between $y_2$ and $y_3$. \\

Interestingly, by rotating the above matrix by the TBM mixing matrix $V_0$, we get :
\begin{eqnarray}
V_0^{T} M_{\nu} V_0 & = & 
\left( \begin{array}{ccc} 
\alpha - \frac{\alpha}{2} \left(2 y_1 - y_2 - y_3 \right) & 0 & 
\frac{\alpha \sqrt{3}}{2} \left(y_2 - y_3 \right) \\
0 & \alpha + 3 \beta - \alpha \left( y_1 + y_2 + y_3 \right) & 0 \\ 
\frac{\alpha \sqrt{3}}{2} \left(y_2 - y_3 \right) & 0 & 
\alpha + \frac{\alpha}{2} \left(2 y_1 - y_2 - y_3 \right) \end{array} \right) ~~.
\end{eqnarray}

As a consequence, the neutrino matrix $M_{\nu}$ is diagonalized by the total unitary matrix $V = V_0 U P_{Maj}$. 
The mixing matrix $U$ is given by : 
\begin{eqnarray}
U & = & \left( \begin{array}{ccc} 
\cos \theta & 0 & e^{- i \delta} \sin \theta \\
0 & 1 & 0 \\
- e^{i \delta} \sin \theta & 0 & \cos \theta \end{array} \right) ~~~~.
\end{eqnarray} 
A straightforward calculation yields that the angle $\theta$ and the CP-phase $\delta$ are : 
\begin{eqnarray}
\tan 2 \theta & = & \frac{\sqrt{X^2 + Y^2}}{Z} \nonumber \\
\tan \delta & = & \frac{Y}{X}
\end{eqnarray}
where 
\begin{eqnarray}
X & = & \sqrt{3} \left( |y_2| \cos \frac{\varphi_2}{2} - |y_3| \cos \frac{\varphi_3}{2} \right) \nonumber \\
Y & = & \sqrt{3} \left( |y_1| |y_2| \sin (\frac{\varphi_1 - \varphi_2}{2}) - 
|y_1| |y_3| \sin (\frac{\varphi_1 - \varphi_3}{2}) + |y_2| |y_3| \sin (\frac{\varphi_2 - \varphi_3}{2}) \right) \nonumber \\
Z & = & 2 |y_1| \cos \frac{\varphi_1}{2} - |y_2| \cos \frac{\varphi_2}{2}-|y_3| \cos \frac{\varphi_3}{2} ~~~~.
\end{eqnarray}
The explicit expression of $V$ is : 
\begin{eqnarray}
V & = & \left( \begin{array}{ccc}
\sqrt{\frac{2}{3}}~\cos \theta & \frac{1}{\sqrt{3}} & \sqrt{\frac{2}{3}}~e^{-i \delta} ~\sin \theta \\
- \frac{\cos \theta}{\sqrt{6}} + \frac{e^{i \delta} ~\sin \theta}{\sqrt{2}} & \frac{1}{\sqrt{3}} 
& - \frac{\cos \theta}{\sqrt{2}} - \frac{e^{-i \delta} ~\sin \theta}{\sqrt{6}} \\
- \frac{\cos \theta}{\sqrt{6}} - \frac{e^{i \delta} ~\sin \theta}{\sqrt{2}} & \frac{1}{\sqrt{3}} 
& \frac{\cos \theta}{\sqrt{2}} - \frac{e^{i \delta} ~\sin \theta}{\sqrt{6}} \end{array} \right) ~.P_{Maj}~~~.
\end{eqnarray}

We immediately obtain the mixing angles :
\begin{eqnarray}
\sin^2 \theta_{13} & = & \frac{2}{3}~ \sin^2 \theta \nonumber \\
\sin^2 \theta_{12} & = & \frac{1}{2 + \cos 2 \theta} \nonumber \\
\sin^2 \theta_{23} & = & \frac{1}{2} \left( 1 + \frac{\sqrt{3} \sin 2 \theta \cos \delta}{2 + \cos 2 \theta} \right) 
\end{eqnarray}
which for $|y_3| = |y_2| = 0$ ( i.e. $\theta_{13}=0$) give the TBM mixing angles. \\

The first and second equation in (20) show that the solar and reactor neutrino mixing angles are related by : 
\begin{eqnarray}
\sin^2 \theta_{12} & = & \frac{1}{3 \cos^2 \theta_{13}} ~~~~.
\end{eqnarray}

Next by considering the third equation in (20), a simple relation between the CP-phase $\delta$ and the mixing angles can be derived. The result for $\cos \delta$ reads : 
\begin{eqnarray} 
\cos \delta & = & - \frac{1}{\sqrt{3}} ~\frac{\cos 2 \theta_{23}}{\cos \theta_{12} \sqrt{3 \sin^2 \theta_{12} -1}} ~~~~. 
\end{eqnarray}
This, in turn, would imply that the solar mixing angle $\theta_{12}$ has to be, 
\begin{eqnarray}
\sin^2 \theta_{12} & > & \frac{1}{3} 
\end{eqnarray}
which is right on the edge from the global fits to the neutrino oscillation data. Such a constraint could be confirmed or ruled out with a little better data. \\

The strength of CP violation in neutrino oscillations is described by the Jarlskog rephasing invariant parameter. It is given by :
\begin{eqnarray}
J_{CP} & = & Im \left( V_{e 2} V_{\mu 3} V_{e 3}^{\star} V_{\mu 2}^{\star} \right) ~=~ - \frac{1}{6 \sqrt{3}} ~\sin 2 \theta \sin \delta \nonumber \\
& = & - \frac{1}{6 \sqrt{3}} ~\frac{Y}{\sqrt{X^2 + Y^2 + Z^2}}  ~~~~~.
\end{eqnarray}
Similarly, the Jarlskog parameter $J_{CP}$ can be written in terms of the solar $\theta_{12}$ and the atmospheric $\theta_{23}$ mixing angles. \\

It proves convenient to use the ratio $R_{\nu} = \frac{\Delta m^2_{21}}{\Delta m^2_{31}}$, to write the mass squared differences as :
\begin{eqnarray}
\Delta m^2_{21} & = & 2 |\alpha|^2 R_{\nu} \sqrt{X^2 + Y^2 + Z^2} \nonumber \\
\Delta m^2_{31} & = & 2 |\alpha|^2 \sqrt{X^2 + Y^2 + Z^2} \nonumber \\
\Delta m^2_{32} & = & 2 |\alpha|^2 (1 - R_{\nu}) ~\sqrt{X^2 + Y^2 + Z^2} ~~~~~.
\end{eqnarray}

In order to confront the above neutrino matrix with the experimental observations (Table 1), we use the constraints on neutrino parameters at $2 \sigma$ and $3 \sigma$ \cite{valle}.
\begin{table}[!h]  
\begin{center}
\begin{tabular}{|c|c|c|c|} 
\hline
\hline  
Parameter & Best fit  & $2 \sigma$ & $3 \sigma$ \\ 
\hline
$\Delta m_{21}^2 [ 10^{-5} eV^2 ]$ & $7.62$ & $7.27 - 8.01$ & $7.12 - 8.20$ \\ 
\hline
$|\Delta m_{31}^2| [ 10^{-3} eV^2 ]$ & $2.55$ & $2.38 - 2.68$ & $2.31 - 2.74$ \\       
                                   & $2.43$ & $2.29 - 2.58$ & $2.21 - 2.64$ \\ 
\hline
$\sin^2 \theta_{12}$               & $0.320$ & $0.29 - 0.35$ & $0.27 - 0.37$ \\ 
\hline
$\sin^2 \theta_{23}$               & $0.613$    & $0.38 - 0.66$ &   $0.36 - 0.68$    \\
                                   & $0.600$    & $0.39 - 0.65$ &  $0.37 - 0.67$      \\
\hline 
$\sin^2 \theta_{13}$               & $0.0246$ & $0.019 - 0.030$ & $0.017 - 0.033$ \\              
                                   & $0.0250$  & $0.020 - 0.030$  & $0.017 - 0.033$ \\
\hline
$\delta$                           & $0.80 \pi$   &    $0 - 2 \pi$  &   $0 - 2 \pi$             \\  
                                   &  $-0.03 \pi$ &  $0 - 2 \pi$    &   $0 - 2 \pi$           \\                                      
\hline 
\end{tabular}
\caption{Global oscillation analysis with best fit for $\Delta m_{21}^2,\Delta m_{31}^2, \sin^2 \theta_{12}, \sin^2 \theta_{23}, \sin^2 \theta_{13}$ and $\delta$ 
the upper and/or lower corresponds to normal and/or inverted neutrino mass hierarchy. }
\label{table1}  
\end{center}
\end{table}  
~\\
For numerical analysis, we use the $3\sigma$ ranges of neutrino oscillation parameters. The neutrino mass matrix $M_{\nu}$ depends on $|\alpha|, |y_1|, |y_2|, |y_3|$ and the phases $\psi, \varphi_1, \varphi_2$ and $\varphi_3$. \\

Since there are many unknown parameters, we consider a particular set of those parameters and show how the measured values of neutrino experiments can be accommodated in our neutrino mass matrix $M_{\nu}$. For simplicity, we take $\varphi_2 = \varphi_3 = 0$, as inputs. \\

To see how the nonzero value of the phase $\varphi_1$ can lift the degeneracy between $m_1$ and $m_2$, we notice that for $y_3 = y_2 = 0$, 
the mass squared differences become : 
\begin{eqnarray}
\Delta m^2_{21} & = & 4 |\alpha|^2 |y_1| \sin \frac{\psi}{2} \sin \frac{\psi - \varphi_1}{2} \nonumber \\
\Delta m^2_{31} & = &  4 |\alpha|^2 |y_1| \cos \frac{\varphi_1}{2} \nonumber \\
\Delta m^2_{32} & = & 4 |\alpha|^2 |y_1| \cos \frac{\psi}{2} \cos \frac{\psi-\varphi_1}{2} ~~~~~~.
\end{eqnarray}
It is worthwhile to remark that for $\varphi_1 = \psi$, the masses $m_1$ and $m_2$ are degenerate and $|m_3| > |m_2|$. Now to separately obtain a nonzero mass 
squared difference $\Delta m^2_{21}$ which is smaller than $\Delta m^2_{31}$, we introduce a small phase difference $\epsilon_1 = \psi - \varphi_1$ between the 
phases $\psi$ and $\varphi_1$. Such a small phase difference will be responsible for lifting the mass degeneracy between the first and second generation. \\

To see the behavior of the mass eigenvalues with respect to $\psi$ and other observables, we take $\epsilon_1 = 6^{\circ}$ as a typical value 
with $|\alpha| =0.1~eV$, $|y_1| = 5 \times 10^{-2}$, $|y_2| =1.3 \times 10^{-3}$ and $|y_3| =1.2 \times 10^{-3}$. Figure 1 shows the variation of the masses 
$|m_i|$ as a function of the phase $\psi$. One clearly observes that Fig. 1 suggests a normal hierarchical ordering pattern for $\psi > 70^{\circ}$. \\

\begin{figure}[hbtp]
\centering
\epsfxsize=6cm
\centerline{\epsfbox{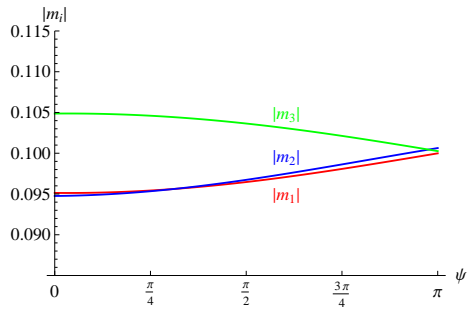}}
\caption{Neutrino's masses versus $\psi$.}
\end{figure} 
Based on the expression of the neutrino mass squared differences, we numerically scan over a broader range of $|y_1|$  $( 0.05 \leq |y_1| \leq 0.4 )$ 
to obtain the restriction on the parameter space of $|y_2|$ and $|y_3|$. We have plotted in Fig. 2 the contour plots of the mass squared 
differences in the two-dimensional parameter spaces $\left( |y_2|, |y_3| \right)$ where we take $\psi = 120^{\circ}, \epsilon_1 =6^{\circ}$ and 
$|\alpha| = 0.1~eV$ as inputs. \\
From Fig. 2, we obtain the allowed ranges of $|y_2|$ and $|y_3|$ for the normal mass hierarchy,
\begin{eqnarray}
|y_i| \leq 0.02~~~~~~~~~\mbox{for}~i=2,3 ~~~~~~~~.
\end{eqnarray}

\begin{figure}[hbtp]
\centering
\epsfxsize=10cm
\centerline{\epsfbox{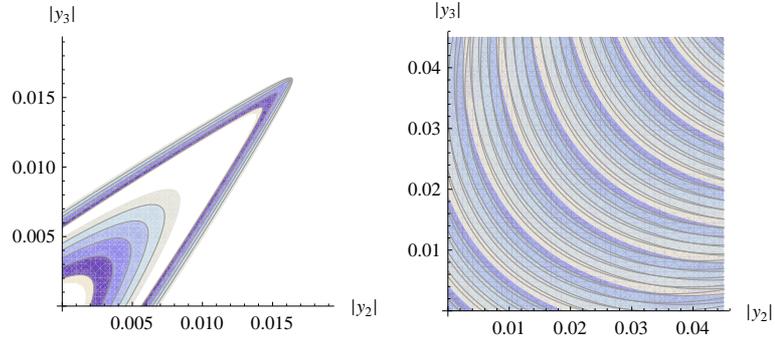}}
\caption{Contour plot of the mass squared differences $\Delta m^2_{21}$ (left panel) and $\Delta m^2_{31}$ (right panel) in the parameter 
space $\left( |y_1|,|y_2| \right)$ .}
\end{figure} 




The ratio $R_{\nu}$ for normal mass hierarchy ordering in the $3 \sigma$ allowed range is : 
\begin{eqnarray}
R_{\nu} & = & \left(2.99^{+0.32}_{-0.34} \right) \times 10^{-2} ~~~~~~.
\end{eqnarray}

Figure 3 shows contour plot of the ratio $R_{\nu}$ in the two parameter space $\left( \psi, \epsilon_1 = \psi -\varphi_1 \right)$ where $|y_1| =0.05, |y_2|=1.3 \times 10^{-3}$ and $|y_3| =1.2 \times 10^{-3}$. As expected, the ratio $R_{\nu}$ at $3 \sigma$ indicates that $\psi$ has to be greater than $70^{\circ}$. \\

\begin{figure}[hbtp]
\centering
\epsfxsize=6cm
\centerline{\epsfbox{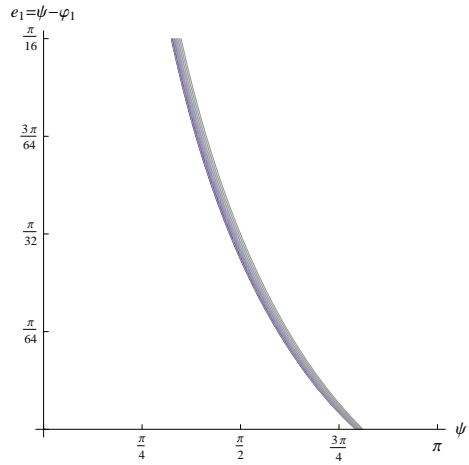}}
  \caption{Contour plot of the ratio $R_{\nu}$ in the parameter space $\left( \psi, \epsilon_1 \right)$.}
\end{figure} 

The departure of the mixing angles from TBM mixing angles depend on the phase $\epsilon_1$ and the two parameters $|y_2|$ and $|y_3|$. Scanning over $|y_2|$ and 
$|y_3|$ within their allowed ranges ( $|y2|, |y_3| \leq 0.02$ ), we investigate how a nonzero $\theta_{13}$ can be obtained for normal mass hierarchy.
As a result of numerical analysis, contour plots in the $\left( \psi, |y_1| \right)$ parameter plane of the mixing angles $\theta_{13}$, $\theta_{12}$ 
and $\theta_{23}$ are shown in Fig. 4 using the experimental constraints on the measured angles. \\
\begin{figure}[hbtp]
\centering
\epsfxsize=10cm
\centerline{\epsfbox{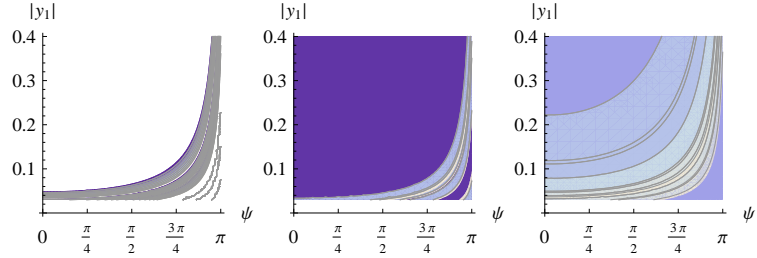}}
  \caption{Contour plots of  reactor angle $\theta_{13}$ (left panel), solar angle $\theta_{12}$ (middle panel ) and atmospheric angle $\theta_{23}$ ( right panel ) 
  in the parameter space $|y_1|$ and $\psi$. }
\end{figure} 




The CP violation parameter $J_{CP}$, which is directly related to the Dirac phase $\delta$, arises due to nonzero value of the difference $|y_3| - |y_2|$ 
in the neutrino matrix $M_{\nu}$. We have plotted in Fig. 5, $J_{CP}$ with respect to $\psi$ using the allowed region of $|y_i$. It leads to values 
of $|J_{CP}|$ around $10^{-3}$. Nonvanishing $|J_{CP}|$ will be explored by the next generation high performance long-baseline neutrino experiments. \\
\begin{figure}[hbtp]
\centering
\epsfxsize=6cm
\centerline{\epsfbox{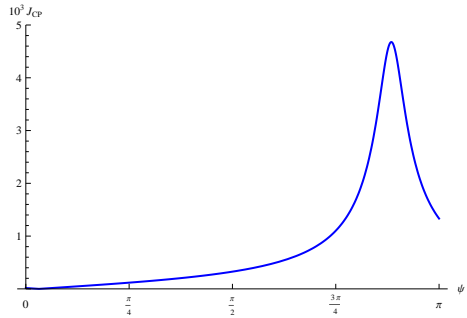}}
  \caption{CP-violating parameter $J_{CP}$ versus $\psi$.}
\end{figure} 

The absolute neutrino mass scale can be probed by nonoscillatory neutrino experiments. Cosmology is sensitive to the sum of neutrino masses 
$\sum |m_i|$. The beta decay endpoint measurements probe the so-called effective electron neutrino mass $m_{\beta}$. The rate of the neutrinoless 
double beta decay depends on the effective Majorana mass of the electron neutrino $m_{\beta \beta}$. \\

Both $m_{\beta \beta}$ and $m_{\beta}$ and the sum of neutrino masses are given by :
\begin{eqnarray}
m_{\beta \beta} & = & |\sum_i m_i V_{e i}^2|  \nonumber \\
m_{\beta} & = & \sqrt{\sum_i m_i^2 |V_{e i}|^2} \nonumber \\
\sum |m_i| & = & |m_1| + |m_2| + |m_3|   ~~~~~~~~~.
\end{eqnarray}

From the neutrino mass $M_{\nu}$, the effective Majorana masses $m_{\beta \beta}$ and $m_{\beta}$ can be written as : 
\begin{eqnarray}
m_{\beta \beta} & = & \frac{|\alpha|}{3}~\sqrt{5 + 4 \cos \psi - 6 |y_1| \left( \sin \frac{\psi -\varphi_1}{2} + 2 \sin \frac{\varphi_1}{2} \right) + 9 |y_1|^2} \nonumber \\
m_{\beta} & = & |\alpha|~\sqrt{1 - \frac{2}{3} y_1 \left( 2 \cos \frac{\varphi_1}{2} + \cos( \psi - \frac{\varphi_1}{2}) \right) +  \frac{4}{3} (y_2+y_3) \cos^2 \frac{\psi}{2} + y_1^2 + y_2^2 +y_3^2} ~~.
\end{eqnarray}
Present cosmological constraints on the sum of neutrino masses $\sum |m_i|$ are in the range $0.44-0.76~eV$ \cite{raffelt}. The Mainz \cite{kraus} and 
Troitsk \cite{troitsk} experiments on the high precision measurement of the end-point part of the $\beta$-spectrum of $^{3}H$ decay found the $95\%$ C.L. 
upper bounds $m_{\beta} \leq 2.3~eV$ ( Mainz ) and $m_{\beta} \leq 2.1~eV$ ( Troitsk). 
Experimental bound on $m_{\beta \beta}$ is below $0.36~eV$ \cite{gomez}. \\
\begin{figure}[hbtp]
\centering
\epsfxsize=14cm
\centerline{\epsfbox{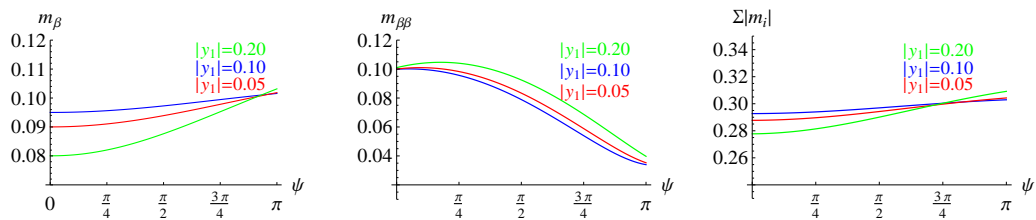}}
  \caption{$m_{\beta}$,$m_{\beta \beta}$ and $\sum |m_i|$ versus $\psi$.}
\end{figure} 

Figure 6 gives the effective electron neutrino mass $m_{\beta}$, the effective Majorana mass $m_{\beta \beta}$ and the sum of neutrino masses $\sum |m_i|$ with respect to $\psi$ for $\epsilon_1 = 6^{\circ}$. It shows the predicted $m_{\beta}$, $m_{\beta \beta}$ and $\sum |m_i|$ are well below the experimental bounds.  
The magnitude of $m_{\beta \beta}$ increases with larger $|y_1|$ values.

\section*{Acknowledgments}
I would like to thank S. Nasri and J. Schechter for reading the manuscript and useful comments.

\end{document}